\def\mevm{~\ensuremath{\mathrm{MeV}/c^2}\/}
\def\mevp{~\ensuremath{\mathrm{MeV}/c}\/}

\def\gevm{~\ensuremath{\mathrm{GeV}/c^2}\/}
\def\gevp{~\ensuremath{\mathrm{GeV}/c}\/}
\def\geve{~\ensuremath{\mathrm{GeV}}}

\def\mbc{\ensuremath{M^{}_{\rm bc}}}
\def\deltaE{\ensuremath{\Delta E}}

\def\klr{\ensuremath{LR}}

\def\ra{\ensuremath{\!\rightarrow\!}}
\def\bbar{\ensuremath{\overline{B}{}^{\,0}}}
\def\qqbar{\ensuremath{q\bar{q}}}
\def\bbbar{\ensuremath{B^0-\bbar}}
\def\bbbar1{\ensuremath{B\bar{B}}}
\def\api{\ensuremath{a_{1}^{\pm}(1260)\ra \rho^{0}\pi^{\pm}}}
\def\rhopi{\ensuremath{\rho^{0}\ra \pi^{+}\pi^{-}}}
\def\a2pi{\ensuremath{B^0\ra a_{2}^{\pm}(1320)\pi^{\mp}}}
\def\ba1pi{\ensuremath{B^0\ra a_{1}^{\pm}(1260)\pi^{\mp}}}
\def\cqq{\ensuremath{q\bar{q}+b\rightarrow c}}
\def\bu{\ensuremath{b\rightarrow u}}
\def\nresrho{\ensuremath{B^0\rightarrow\rho^{0} \pi^{+}\pi^{-}}}
\def\nrespi{\ensuremath{B^0\rightarrow\pi^{+} \pi^{-}\pi^{+}\pi^{-}}}

\documentclass[aps,prl,preprint,tightenlines,superscriptaddress,showpacs,byrevtex]{revtex4}

\usepackage{graphicx} 
\usepackage{dcolumn}  

\graphicspath{{ps}}

\renewcommand{\arraystretch}{1.1}

\begin{document}


\preprint{\vbox{ \hbox{   }
                 \hbox{BELLE-CONF-0543}
                 \hbox{EPS05-518} 
}}

\title{ $\quad$\\[0.5cm]  Observation of  $B^{0}
  \to \mathrm{a}_{1}^{\pm}(1260)\pi^{\mp}$ }



\affiliation{Aomori University, Aomori}
\affiliation{Budker Institute of Nuclear Physics, Novosibirsk}
\affiliation{Chiba University, Chiba}
\affiliation{Chonnam National University, Kwangju}
\affiliation{University of Cincinnati, Cincinnati, Ohio 45221}
\affiliation{University of Frankfurt, Frankfurt}
\affiliation{Gyeongsang National University, Chinju}
\affiliation{University of Hawaii, Honolulu, Hawaii 96822}
\affiliation{High Energy Accelerator Research Organization (KEK), Tsukuba}
\affiliation{Hiroshima Institute of Technology, Hiroshima}
\affiliation{Institute of High Energy Physics, Chinese Academy of Sciences, Beijing}
\affiliation{Institute of High Energy Physics, Vienna}
\affiliation{Institute for Theoretical and Experimental Physics, Moscow}
\affiliation{J. Stefan Institute, Ljubljana}
\affiliation{Kanagawa University, Yokohama}
\affiliation{Korea University, Seoul}
\affiliation{Kyoto University, Kyoto}
\affiliation{Kyungpook National University, Taegu}
\affiliation{Swiss Federal Institute of Technology of Lausanne, EPFL, Lausanne}
\affiliation{University of Ljubljana, Ljubljana}
\affiliation{University of Maribor, Maribor}
\affiliation{University of Melbourne, Victoria}
\affiliation{Nagoya University, Nagoya}
\affiliation{Nara Women's University, Nara}
\affiliation{National Central University, Chung-li}
\affiliation{National Kaohsiung Normal University, Kaohsiung}
\affiliation{National United University, Miao Li}
\affiliation{Department of Physics, National Taiwan University, Taipei}
\affiliation{H. Niewodniczanski Institute of Nuclear Physics, Krakow}
\affiliation{Nippon Dental University, Niigata}
\affiliation{Niigata University, Niigata}
\affiliation{Nova Gorica Polytechnic, Nova Gorica}
\affiliation{Osaka City University, Osaka}
\affiliation{Osaka University, Osaka}
\affiliation{Panjab University, Chandigarh}
\affiliation{Peking University, Beijing}
\affiliation{Princeton University, Princeton, New Jersey 08544}
\affiliation{RIKEN BNL Research Center, Upton, New York 11973}
\affiliation{Saga University, Saga}
\affiliation{University of Science and Technology of China, Hefei}
\affiliation{Seoul National University, Seoul}
\affiliation{Shinshu University, Nagano}
\affiliation{Sungkyunkwan University, Suwon}
\affiliation{University of Sydney, Sydney NSW}
\affiliation{Tata Institute of Fundamental Research, Bombay}
\affiliation{Toho University, Funabashi}
\affiliation{Tohoku Gakuin University, Tagajo}
\affiliation{Tohoku University, Sendai}
\affiliation{Department of Physics, University of Tokyo, Tokyo}
\affiliation{Tokyo Institute of Technology, Tokyo}
\affiliation{Tokyo Metropolitan University, Tokyo}
\affiliation{Tokyo University of Agriculture and Technology, Tokyo}
\affiliation{Toyama National College of Maritime Technology, Toyama}
\affiliation{University of Tsukuba, Tsukuba}
\affiliation{Utkal University, Bhubaneswer}
\affiliation{Virginia Polytechnic Institute and State University, Blacksburg, Virginia 24061}
\affiliation{Yonsei University, Seoul}
  \author{K.~Abe}\affiliation{High Energy Accelerator Research Organization (KEK), Tsukuba} 
  \author{K.~Abe}\affiliation{Tohoku Gakuin University, Tagajo} 
  \author{I.~Adachi}\affiliation{High Energy Accelerator Research Organization (KEK), Tsukuba} 
  \author{H.~Aihara}\affiliation{Department of Physics, University of Tokyo, Tokyo} 
  \author{K.~Aoki}\affiliation{Nagoya University, Nagoya} 
  \author{K.~Arinstein}\affiliation{Budker Institute of Nuclear Physics, Novosibirsk} 
  \author{Y.~Asano}\affiliation{University of Tsukuba, Tsukuba} 
  \author{T.~Aso}\affiliation{Toyama National College of Maritime Technology, Toyama} 
  \author{V.~Aulchenko}\affiliation{Budker Institute of Nuclear Physics, Novosibirsk} 
  \author{T.~Aushev}\affiliation{Institute for Theoretical and Experimental Physics, Moscow} 
% ...
\collaboration{The Belle Collaboration}
\noaffiliation

\noaffiliation

\begin{abstract}
We report a new measurement of the branching fraction
of neutral $B$
meson decays  \ba1pi\ with  \api\ and \rhopi\ 
using $275\times 10^{6}$ \bbbar1\ pairs 
collected by the Belle detector at the KEKB 
asymmetric-energy $e^{+}e^{-}$
collider. We measure the
branching fraction  $\mathcal{B}(\ba1pi)=(48.6\pm 4.1(stat)\pm 3.9(syst)) \times 10^{-6}$. Using a relativistic Breit-Wigner parameterization,
we measure the mass and width of the
$a_{1}(1260)$ to be   $m_{a_1}=1197\pm 34 \mevm$ and 
$\Gamma_{a_1}=305\pm 43 \mevm$, respectively. \\

\end{abstract}


\maketitle



The Kobayashi and Maskawa (KM) model explains the source
of $CP$ violation in terms of a single  complex phase in the quark
mixing matrix~\cite{kobayashi}. Measurements of the $CP$ violating
asymmetry parameter $\sin2\phi_1$ by the Belle~\cite{abe} and 
BaBar~\cite{aub} collaborations have established $CP$ violation in the neutral
 $B$ meson system. 
 Measurements of other $CP$ violating asymmetry parameters provide
 important tests of the KM model.  Decay modes where the 
 $b\rightarrow u\bar{u}d$ contribution is dominant, such as 
$B^{0}\rightarrow\pi^{+}\pi^{-}$, $B^{0}\rightarrow\rho\pi$, 
$B^{0}\rightarrow\rho\rho$  and 
$B^{0}\rightarrow a_{1}\pi$  can be used to measure the 
Cabibbo-Kobayashi-Maskawa  angle    
$\phi_2$ (also denoted $\alpha)$~\cite{aleksan}.
Unlike  $B^{0}\rightarrow\pi^{+}\pi^{-}$ decay, 
$B^{0}\rightarrow a_{1}^{+}\pi^{-}$ 
 decay
is not a $CP$ eigenstate, and four flavor-charge configurations $(B^{0}(\bar{B^0})\rightarrow a_{1}^{\pm}\pi^{\mp})$ must be considered to measure the angle 
$\phi_{2}$. 

The CLEO  collaboration has quoted
an upper limit of $49\times 10^{-5}$ at the 90\%\
C.L.  for the branching fraction of \ba1pi~\cite{cleo}, while
the DELPHI collaboration has given
an  upper limit of
$28\times 10^{-5}$ for the branching fraction of $B^{0}\rightarrow 4\pi$
at the  90\%\ C.L.~\cite{delphi}.
Recently the BaBar collaboration has reported a measured branching fraction of 
 $\mathcal{B}(\ba1pi)=(40.2\pm 3.9(stat)\pm 3.9(syst)) \times 10^{-6}$~\cite{bab2}. In some previous studies   discrepancies 
have been found in the $a_{1}(1260)$ properties. 
Among these studies, the
 analyses involving hadronic events~\cite{pernegr} and $\tau$ 
decays~\cite{asner} are important.
Therefore, it is important to measure the branching fraction of
this decay mode. A new precise measurement of the  $a_{1}(1260)$ properties
is also useful.  

We present in this paper a measurement of the branching fraction of
\ba1pi\ with \api\ and \rhopi~\cite{b0bar}. In this analysis, 
we assume that the main contributions come from
the \ba1pi\  and ${B^0\ra a_{2}^{\pm}(1320)\pi^{\mp}}$
decays, and neglect any  interferences between 
decay modes   consisting of  four charged pions in the final
state. We also assume that $a_{1}(1260)$ decays to 
$\rho\pi$ only.


The analysis is  based on a  253~fb$^{-1}$ data sample
containing  $275$ million \bbbar1\  pairs. The 
data were collected
with the Belle detector~\cite{Belle} at the KEKB asymmetric-energy 
$e^{+}e^{-}$ (3.5 on 8 GeV) collider~\cite{kurokawa} operating
at the $\Upsilon(4S)$ resonance ($\sqrt{s}=10.58$~GeV) with a peak
luminosity that exceeds $1.5\times10^{34}~{\rm cm}^{-2}{\rm s}^{-1}$.
The Belle detector is a large-solid-angle magnetic
spectrometer that
consists of a silicon vertex detector (SVD),
a 50-layer central drift chamber (CDC), an array of
aerogel threshold \v{C}erenkov counters (ACC), 
a barrel-like arrangement of time-of-flight
scintillation counters (TOF) and an electromagnetic calorimeter
comprised of CsI(Tl) crystals  located inside 
a super-conducting solenoid coil that provides a 1.5~T
magnetic field.  An iron flux-return located outside 
the coil is instrumented to detect $K_L^0$ mesons and to identify
muons.  
Two inner detector configurations were used. A 2.0 cm radius beampipe
and a 3-layer silicon vertex detector were used for the first sample
of 152 million $B\bar{B}$ pairs, while the remaining 123 million 
$B\bar{B}$ pairs were recorded using a 1.5
cm radius beam pipe, a 4-layer silicon detector and a small-cell inner
drift chamber~\cite{Ushiroda}.  

For the present analysis several Monte-Carlo (MC) samples 
(signal and backgrounds)
are generated with 
EvtGen~\cite{Lange}. The Belle detector response is simulated with a
GEANT3-based program~\cite{geant}. 


We reconstruct the $B^{0}\rightarrow a_{1}^{\pm}(1260)\pi^{\mp}$ 
candidate using four
charged tracks originating  from the beam interaction region and each having a 
momentum transverse to the beam, $p_{t}$, greater than $100\mevp$.  To identify  $K$ and $\pi$ mesons, we form a
 $K(\pi)$ likelihood $L_{K}(L_{\pi})$
by combining information from the CDC $(dE/dx)$, the TOF and the ACC. Discrimination between pions and kaons is achieved through the likelihood ratio $R(K/\pi)=L_{K}/(L_{\pi}+L_{K})$. 
Charged tracks with $R(K/\pi)<0.4$ are regarded as pions.
Positively identified electrons and muons  are rejected.

In the event selection we reconstruct each $\rho^{0}$ candidate by
combining 
a $\pi^{+}$ and a $\pi^{-}$ such that the invariant mass of the 
pion pair ($m_{\pi^{+}\pi^{-}}$) satisfies
$0.55\gevm<m_{\pi^{+}\pi^{-}}<1.15\gevm$.
We reconstruct the $a^{\pm}_{1}$(1260) candidates by combining
another charged pion  with the reconstructed  $\rho^{0}$.
The $a^{\pm}_{1}$(1260) candidates are then selected 
on the invariant mass: $0.80\gevm<m_{a_{1}}<1.775\gevm$. 
 
Candidate $B$  mesons are identified using three kinematic 
variables: the beam-energy-constrained mass, $M_{\rm bc}=\sqrt{E_{\rm beam}^{2} - p_B^{2}}$,
the energy difference, $\deltaE=E_{B} - E_{\rm beam}$, and the helicity
angle  of the $a_{1}(1260)$ meson.
Here, $E_{B}$ and $p_{B}$ are the reconstructed 
energy and momentum of the $B$ candidate in the center of mass (c.m.)
frame, and  $E_{\rm beam}$ is the beam energy in the c.m. frame. 
The  angle between the  $\rho^{0}$ momentum vector 
and the direction opposite to the $B^{0}$ in the $a_{1}^{\pm}(1260)$
rest frame is defined as the  helicity angle $(\theta_{\mathrm{hel}})$.
To extract the signal yield we select events in the region 
$5.21\gevm< \mbc <5.29\gevm$, $-0.10\geve < \Delta E < 0.12\geve$ and
$-1.0\!<\! \cos\theta_{\mathrm{hel}} \!<\!1.0$ (loose signal region).
The momentum of the bachelor pion (the pion that comes directly from the 
decay of \ba1pi) in   the $B^{0}$ rest 
frame is required to satisfy  $2.2\gevp< p_{\mathrm{bach}} <2.7\gevp$.
We find 39\%\ of the reconstructed events have more
than one $B$ candidate after all selection requirements. 
We select the best $B$ meson 
candidate that has the minimum 
$\chi^{2}$ value in the  vertex fit to the four charged pion tracks.

We find 14\%\ of signal events in the MC simulation 
are incorrectly reconstructed by including at least one charged track
from the other  $B$  meson.
The remaining 86\%\ of signal events are reconstructed from the
correct set of final state pions. However there remains the
possibility of incorrectly assigning the final state pions to the
intermediate mesons in the decay. 
We find   17\%\ of events have multiple allowable pion combinations.
If multiple combinations
exist which satisfy the invariant mass constraint on the  $\rho^{0}$
and $a_{1}^{\pm}(1260)$ mesons, we choose the combination such
that the daughter pions of the $\rho^{0}$ have the larger 
transverse momenta.
With this selection,
the probability of selecting the wrong $\rho^{0}$ and 
$a_{1}^{\pm}(1260)$  combination
is found to be 27\%\ in the signal MC simulation.

The dominant background in the \ba1pi\ candidates   comes from the
continuum $e^+e^-\ra q\bar{q}$ $(q=u,d,s,c)$ processes, since the final states 
contain only charged pions. To suppress the continuum
background  we apply a Likelihood Ratio (\klr) cut. 
The correlated event shape variables are put into a Fisher
discriminant~\cite{fish}
to form a single variable, which is combined with the
cosine of the angle between the $B$ meson flight direction and the beam axis,
to form a LR. 
We determine the  \klr\ cut value of  $LR>0.80$ by optimizing the figure-of-merit as a function of the \klr.
From MC studies we find that this cut ($LR>0.80$) reduces
the continuum and  $b\ra c$ generic decay
backgrounds by 97.4\%\ and 76.6\%, respectively.

We estimate the background contribution
from rare $b\ra u$  decays 
using large (2500~fb$^{-1}$ equivalent) MC samples. 
Since the  \a2pi\ decay also has a
 four pion final state, a significant background contribution is expected from this
decay mode. 
The $a_{1}(1260)$ meson has a mass of $1.230$\gevm\ and a width of 
 $0.25$-$0.60$\gevm, whereas the $a_{2}(1320)$ has a mass of $1.318$\gevm\ and a width of 0.107\gevm~\cite{eidleman}. The upper limit for the branching
 fraction  of $B^0\ra a_{2}^{\pm}(1320)\pi^{\mp}$  is given as $3.0\times 10^{-4}$ (90\%\ C.L.)~\cite{eidleman}. 
From MC studies, we find  that the  $a_{2}(1320)$ has a large overlap with the 
$a_{1}(1260)$ in the $m_{3\pi}$ distribution, where $m_{3\pi}$  is the
invariant mass of three pions.
 However, since the spins of 
 $a_{1}(1260)$ and  $a_{2}(1320)$ are different,
 the helicity angle distributions for the $a_{1}(1260)$
and  $a_{2}(1320)$ mesons differ. A three dimensional (3D)
 (\mbc-\deltaE-$\cos\theta_{\mathrm{hel}}$)  fit to the data is used to discriminate the signal
from the  $a_{2}(1320)$  background.
We note that this method is nearly independent of assumptions about the masses and widths of the $a_{1}(1260)$  and   $a_{2}(1320)$ mesons.



The signal yields are extracted from a 3D (\mbc-\deltaE-$\cos\theta_{\mathrm{hel}}$)
unbinned maximum likelihood (ML) fit to the data.
We fit to the candidates  in the loose signal region.
We define the likelihood function as:
\begin{eqnarray}
{\cal L} & = & \prod_i\ \left[\,
f^{}_{a_{1}\pi}\,(P^{}_{a_{1}\pi}(M^{}_{{\rm bc}\,i}).P^{}_{a_{1}\pi}(\Delta E^{}_i).P^{}_{a_{1}\pi}(\cos\theta^{}_{{\rm hel}\,i})) \right. \nonumber\\ 
& & \hskip0.2in \left. \ +\ 
f^{}_{a_{2}\pi}\,(P^{}_{a_{2}\pi} 
(M^{}_{{\rm bc}\,i}).P^{}_{a_{2}\pi}(\Delta E^{}_i).P^{}_{a_{2}\pi}(\cos\theta^{}_{{\rm hel}\,i})) \right. \nonumber\\
& & \hskip 0.2in \left. \ +\
f^{}_{b\rightarrow u}\,(P^{}_{b\rightarrow u} (M^{}_{{\rm bc}\,i}).P^{}_{b\rightarrow u}(\Delta E^{}_i).P^{}_{b\rightarrow u}(\cos\theta^{}_{{\rm hel}\,i})) \right.  \nonumber\\ 
 & & \hskip 0.2in \left. \ +\ 
f^{}_{(q\bar{q}+b\rightarrow c)}\,(P^{}_{(q\bar{q}+b\rightarrow c)}(M^{}_{{\rm bc}\,i}).P^{}_{(q\bar{q}+b\rightarrow c)}(\Delta E^{}_i).P^{}_{(q\bar{q}+b\rightarrow c)}(\cos\theta^{}_{{\rm hel}\,i}))],  \right.
\end{eqnarray}
where $i$ runs over all events in the sample.
The coefficients  $f^{}_{a_{1}\pi}$,
$f^{}_{a_{2}\pi}$,
$f^{}_{b\rightarrow u}$ and 
$f^{}_{(q\bar{q}+b\rightarrow c)}$
denote the fractions of events from \ba1pi\ signal, \a2pi\ background,
charmless rare $b\ra u$ decay background   and continuum and generic
decay (\qqbar\ +$b\ra c$) background, respectively. 

The $P(M^{}_{{\rm bc}\,i})$, $P(\Delta E^{}_i)$ and $P(\cos\theta^{}_{{\rm hel}\,i})$
are the Probability Density Functions (PDFs) 
in \mbc, \deltaE\  and $\cos\theta^{}_{{\rm hel}}$ for the signal and
background contributions, respectively. 
The PDF for the \ba1pi\ signal is taken from a MC simulation,
with the calibration correction obtained from the
$B^0\ra D^-\pi^+, D^-\ra K^+\pi^-\pi^-$ control sample.
The \mbc\ and \deltaE\ signal PDFs are described with a double Gaussian function and the  $\cos\theta^{}_{{\rm hel}}$ signal PDF with a  third order polynomial function. The PDFs for continuum and $b\rightarrow c$ decays are taken from the sideband
data, where the  \mbc\ sideband region is defined as
$ 5.21\gevm<\!M^{}_{\rm bc\!}\!<\!5.26\gevm$, 
 and the
\deltaE\  sideband region is defined as   
$0.10\geve\!<\!\Delta E\!<\!0.35\geve$. 
The shapes of the \mbc, \deltaE\ and  $\cos\theta^{}_{{\rm hel}}$  distributions for the sideband
data are assumed to be the same  for continuum  and 
generic $b\ra c$ backgrounds. This assumption is justified using a MC 
simulation.
The \mbc, \deltaE\  and   $\cos\theta^{}_{{\rm hel}}$ PDFs for continuum  and 
generic $b\ra c$ 
backgrounds are 
an ARGUS function, a second order
polynomial and a  third order polynomial, respectively.
The PDF for \a2pi\ background
is also taken from a MC simulation. The \mbc\ and \deltaE\
shapes are modelled as a single Gaussian and the 
 $\cos\theta^{}_{{\rm hel}}$ distribution  is described by a 
 4th order polynomial function.  
The $b\ra u$ background  PDF is taken from
 MC samples.

\begin{figure}[htb]
\includegraphics[width=0.52\textwidth]{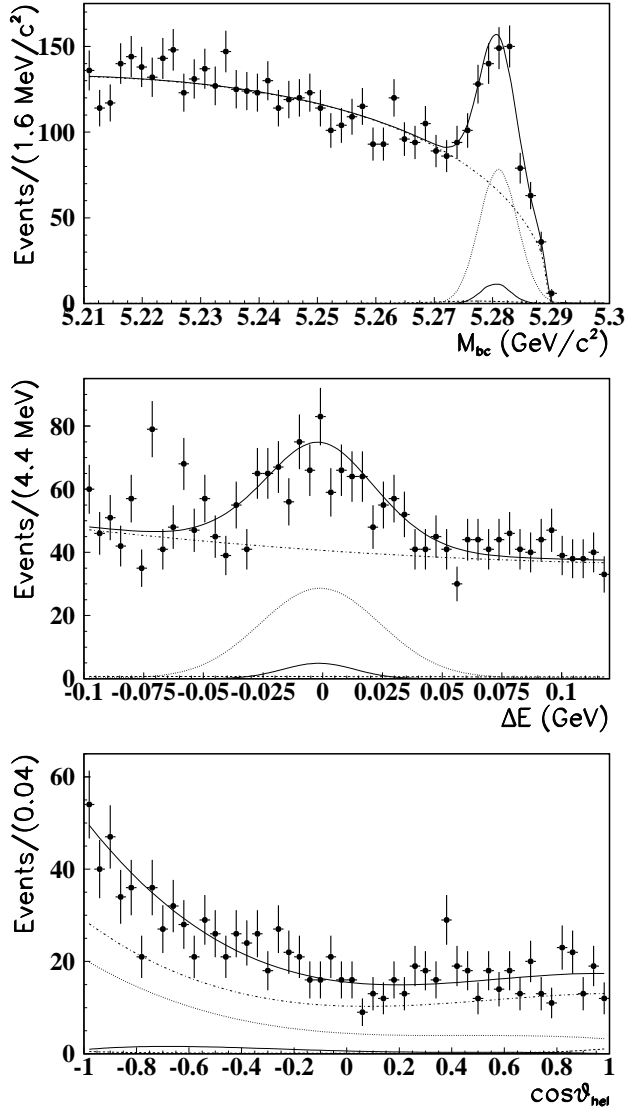}
\caption{  \mbc, \deltaE\  and  $\cos\theta^{}_{{\rm hel}}$ projections for the
 3D unbinned ML fit for the selected sample.  
The top, middle and bottom figures show the \mbc,
 \deltaE\  and  $\cos\theta^{}_{{\rm hel}}$ distributions, respectively.
 In each plot, the top solid line, the dotted
dash line, the dotted line, the bottom solid line and the bottom
dash line represent the total fit function, the continuum background,
the signal yield, the \a2pi\ background and the rare decay background,
respectively.
}
\label{fig:projections1}
\end{figure}

In the fit we float the
$f^{}_{ a_{1}\pi }$, $f^{}_{ a_{2}\pi }$ and
$f^{}_{b\rightarrow u}$ coefficients and constrain $f^{}_{(q\bar{q}+b\rightarrow c)}$ according to 
$f^{}_{(q\bar{q}+b\rightarrow c)}= 1-(f^{}_{ a_{1}\pi }
+f^{}_{ a_{2}\pi}+f^{}_{b\rightarrow u})$.
Figure~\ref{fig:projections1}  shows the
projections of the \mbc, \deltaE\ and  $\cos\theta^{}_{{\rm hel}}$ distribution.
The \mbc\ plot  shows the projection with
$-0.04\geve\!<\!\Delta E\!<\!0.04\geve$, 
the \deltaE\ plot  shows the projection with
$5.27\gevm\!<M^{}_{\rm bc}\!<\!5.29\gevm$ 
and the $\cos\theta^{}_{{\rm hel}}$
plot shows the projection with 
$5.27\gevm\!<M^{}_{\rm bc}\!<\!5.29\gevm$ and $-0.04\geve\!<\!\Delta E\!<\!0.04\geve$ (tight signal region).

A total of 13541 events survive after all the selection requirements and are included
in the fit.  From the fit we find
 $394\pm 33$ \ba1pi\ signal events, $34\pm 38$
\a2pi\ background events and 
$26\pm 30$ rare decay background events. 
The remaining events in the fit are from  continuum and generic  $b\ra c$ 
background.

We calculate the branching fraction using the following equation:

\begin{eqnarray}
\mathcal{B} (\ba1pi) & = & \frac{N_{a_{1}^{\pm}\pi^{\mp}}} 
{ N_{\bbbar1} \cdot \varepsilon \cdot Br^{sub} \cdot \varepsilon_{PID}},   
\end{eqnarray}
where $N_{a_{1}^{\pm}\pi^{\mp}}$ is the number of \ba1pi\ events obtained from
the \mbc-\deltaE-$\cos\theta^{}_{{\rm hel}}$ simultaneous fit, $N_{\bbbar1}$ is the
number of \bbbar1\  pairs,  $Br^{sub}$ is the branching fraction of the $a_{1}^{\pm}(1260)$
decay, $\varepsilon$ is the reconstruction 
efficiency estimated from MC, and $\varepsilon_{PID}$ is the 
correction for the difference in particle identification (PID) efficiency
for pions between data and a MC simulation. The estimated values for 
 $\varepsilon$ and $\varepsilon_{PID}$ are $0.061$ and
$0.962$, respectively. The branching fraction
of   $a_{1}^{\pm}(1260)\ra \rho^{0}\pi^{\pm}$ decay is assumed to be 0.5. 

Inserting all these values  in Eq.~(2) we measure the branching fraction 
$\mathcal{B}(\ba1pi)=(48.6\pm 4.1(stat))\times 10^{-6}$.

\begin{figure}[htb]
\includegraphics[width=0.49\textwidth]{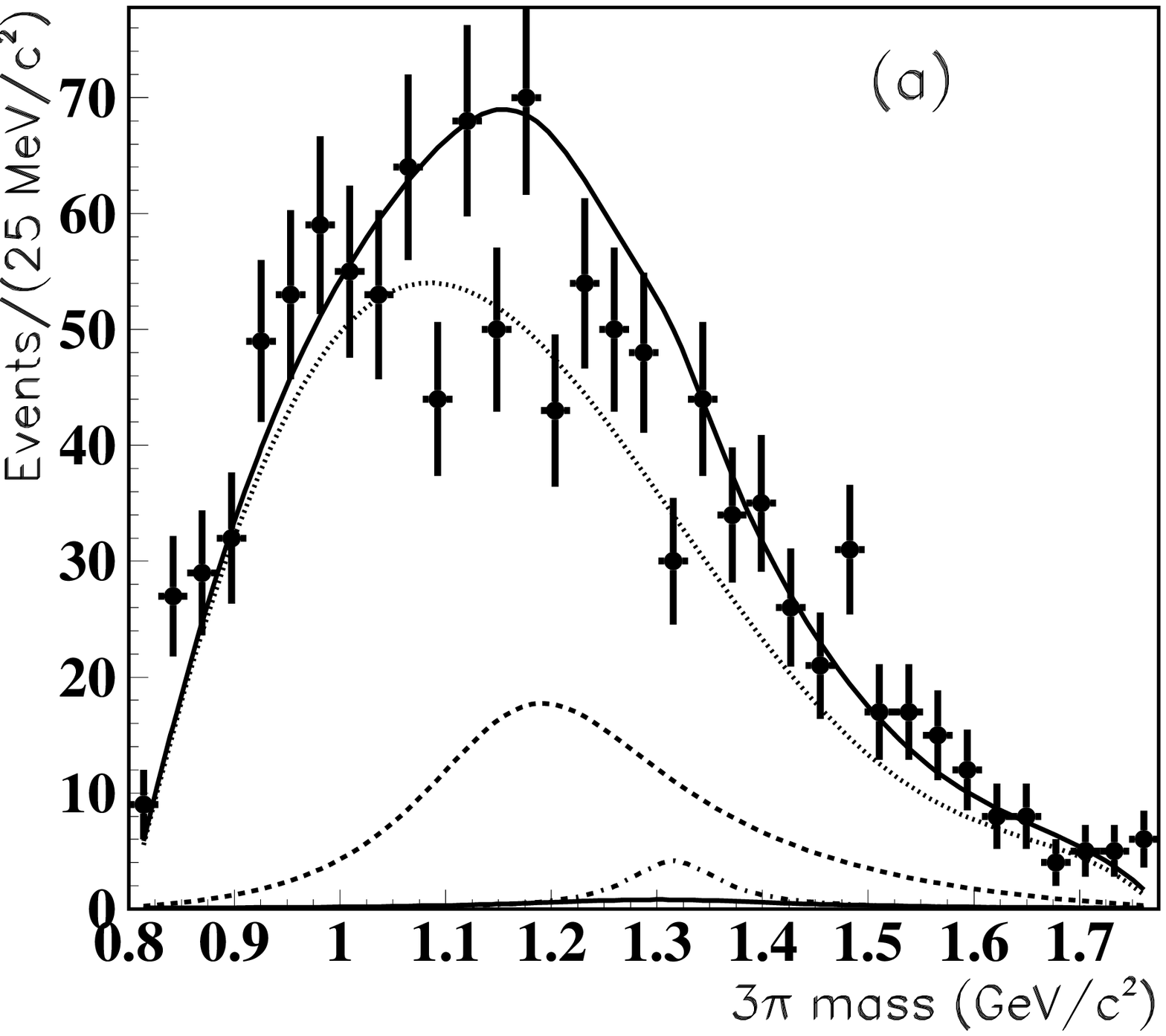}
\includegraphics[width=0.49\textwidth]{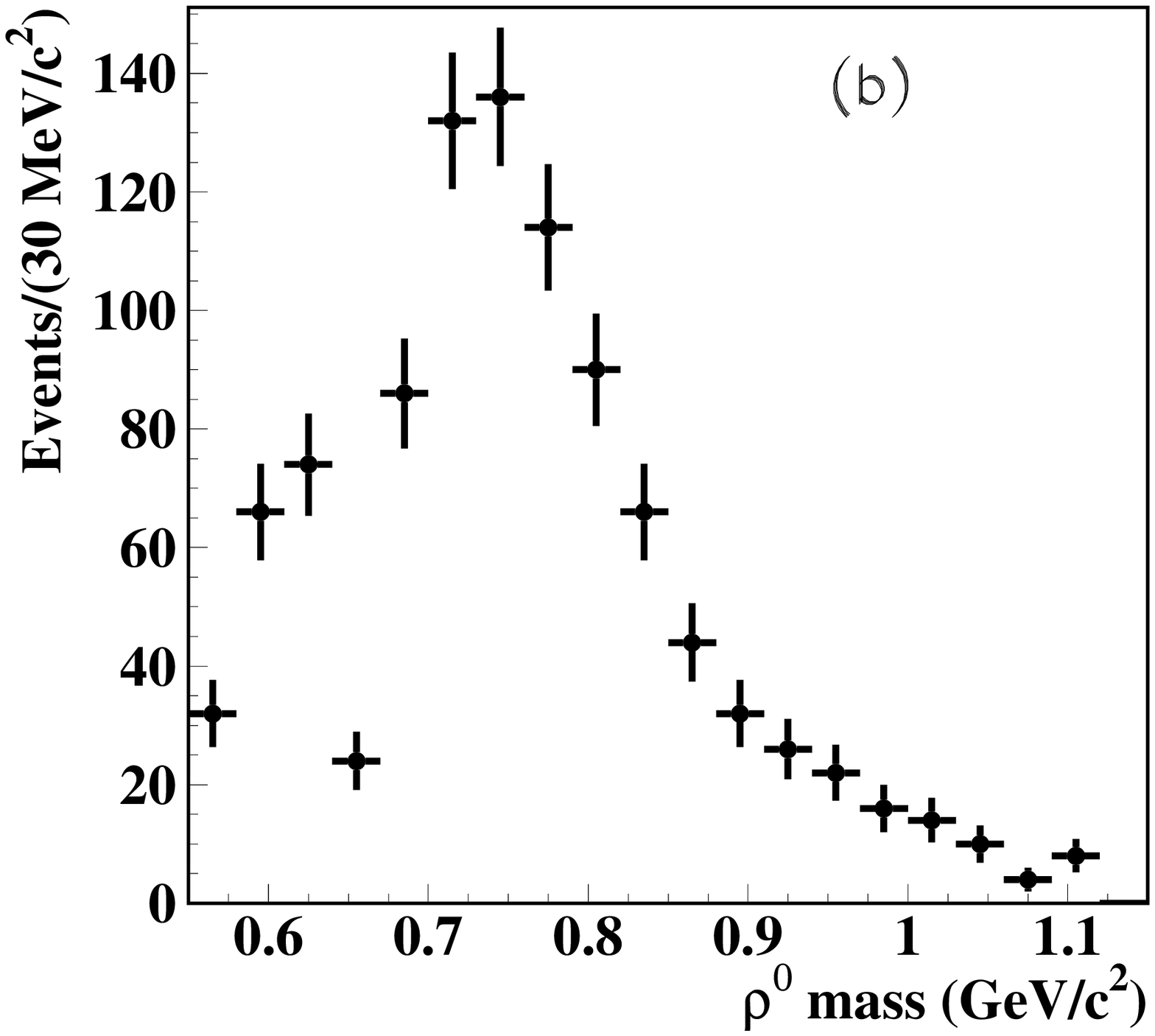}
\caption{ (a) $m_{3\pi}$  and (b) $\rho^{0}$ mass  
distributions for the events in the 
tight signal region.   
In plot (a) the top solid line shows the total fit, the dotted line shows
the contribution from \cqq, the dashed line shows the contribution from
\ba1pi, the dotted-dashed line shows the contribution from \a2pi\ and
the bottom solid line represents the contribution from the \bu\ background.
}
\label{fig:a1mass_data1}
\end{figure}

To measure the  $a_{1}(1260)$ properties (mass and width)
and the contribution from the non-resonant components,
we select $1193$ \ba1pi\ candidates with the tight signal region
to enhance 
signal events.
We fit
the  $a_{1}(1260)$ mass within the mass window
$0.8\gevm\!<\!m_{3\pi}\!<\!1.775\gevm$.
For the $m_{3\pi}$  fit, we use the following likelihood function:

\begin{eqnarray}
{\cal L} & = & \prod_i\ \left[\,
f^{}_{a_{1}\pi}\,P^{}_{a_{1}\pi}(m^{i}_{3\pi})\ +\
f^{}_{a_{2}\pi}\,P^{}_{a_{2}\pi}(m^{i}_{3\pi})\ +\
\right. \nonumber\\  & & \nonumber\\
 & & \hskip0.30in \left.
f^{}_{b\rightarrow u}\,P^{}_{b\rightarrow u} (m^{i}_{3\pi})\ +\ 
f^{}_{(q\bar{q}+b\rightarrow c)}\,P^{}_{(q\bar{q}+b\rightarrow c)}(m^{i}_{3\pi})\,\right],
\end{eqnarray}
where $i$ runs over all events. The coefficients
$f_{a_{1}\pi}$, $f_{a_{2}\pi}$, $f^{}_{b\rightarrow u}$, 
and $f^{}_{(q\bar{q}\,+\,b\rightarrow c)}$ represent 
fractions  of events from signal, \a2pi\ background,
$b\ra u$ background  and 
${(q\bar{q}\,+\,b\rightarrow c)}$ background, respectively. 
The $P(m^{}_{3\pi})$ are the PDFs for the corresponding signal and background contributions.
In the present analysis we impose several kinematical cuts in the event
selection, which produce a non-uniform efficiency  in $m_{3\pi}$. 
We obtain an  efficiency correction curve from a MC simulation, which is 
applied to the  relativistic  Breit-Wigner
(RBW)  function for the 
$a_{1}(1260)$ and  $a_{2}(1320)$ PDFs.
The sideband events and MC are  used to determine
the  background shapes for 
${(q\bar{q}\,+\,b\rightarrow c)}$ and \bu, respectively.
The  PDFs for  the (\cqq) and \bu\ backgrounds
are  5th and 6th order polynomial functions, respectively.
In the fit we fix the fractions of \ba1pi, \a2pi\ and \bu\ 
to those
obtained  from the
3D \mbc-\deltaE-$\cos\theta^{}_{{\rm hel}}$ fit \footnote[2]{From
3D(\mbc-\deltaE-$\cos\theta^{}_{{\rm hel}}$) fit the fractions for the signal and backgrounds are measured. Here we use those fractions with an  efficiency correction for the 
tight signal region cut of $m_{a_{1}}$  selection ranges.}. We allow the
mean and width  of the $a_{1}^{\pm}(1260)$ RBW function to float in the fit,
while those of   $a_{2}^{\pm}(1320)$ are fixed to the world 
averages~\cite{eidleman}.
Figure~\ref{fig:a1mass_data1} shows the $m_{3\pi}$ and $\rho^{0}$ mass 
distributions for the events that
survive after the  tight signal region cut.
From the fit we determine the mass and width of the $a_{1}(1260)$ 
as $1197\pm 34(stat)\mevm$ and $305\pm 43(stat)\mevm$, respectively.

In order to measure the non-resonant components,
we extend the  likelihood function of Eq.~(3)
by adding non-resonant contributions from
$B^0\ra\rho^0\pi^+\pi^-$ and $B^0\ra\pi^+\pi^-\pi^+\pi^-$. 
The extended likelihood function includes the additional coefficients 
$f^{}_{\rho\pi\pi}$ and $f^{}_{4\pi}$ which
represent the fractions of the non-resonant
$B^0\ra\rho^0\pi^+\pi^-$ and
$B^0\ra\pi^+\pi^-\pi^+\pi^-$ decays, respectively. 
The PDFs for $B^0\ra\rho^0\pi^+\pi^-$ and
$B^0\ra\pi^+\pi^-\pi^+\pi^-$ decays are taken from MC.
We perform the  $m_{3\pi}$   fit
in the $0.80\gevm< m^{}_{3\pi}<1.775\gevm$  range to extract
the non-resonant components. 
In the fit we fix the  \bu\ and  $(q\bar{q}+b\rightarrow c)$ fractions to 
those obtained from the
\mbc-\deltaE-$\cos\theta^{}_{{\rm hel}}$ fit. We also fix the
 values for the mean and width
of the $a_{1}(1260)$ to $1230$\mevm\ and $300$\mevm, respectively,
which lies in the world  average limits~\cite{eidleman}. 
We  apply the constraint: $1-f^{}_{b\rightarrow u} - f^{}_{(q\bar{q}+b\rightarrow c)}=f^{}_{a_{1}\pi}+f^{}_{a_{2}\pi} +f^{}_{\rho\pi\pi}+f^{}_{4\pi}$. Thus
there are four free parameters in the fit: 
$f^{}_{a_{1}\pi}$, $f^{}_{a_{2}\pi}$, $f^{}_{\rho\pi\pi}$ and $f^{}_{4\pi}$. 
We combine
the $f^{}_{\rho\pi\pi}$ and $f^{}_{4\pi}$
fractions and estimate the total non-resonant component from
\nresrho\ and \nrespi. We obtain a 
non-resonant component fraction of $5.3\pm 5.4$\%,
which is  consistent with 0. Accordingly, we include the non-resonant contribution as a systematic
error in the branching fraction measurement.

The measured signal yield from the 3D fit  contains
 several systematic uncertainties.
The systematic error in the signal yield is estimated by varying each 
parameter of the fit by $\pm1\sigma$ from the nominal values.
The shifts in the signal yields are then added in quadrature.
The corresponding uncertainties are listed in Table~\ref{tab:syst_meth2}.
We assign the uncertainty in the  track reconstruction efficiency to be $\pm 4.8$\%. We measure
the uncertainty in the efficiency of the PID selection of the charged pions to be  $\pm 1.4$\%. A systematic error of 
 $-5.3$\%\ is assigned to a potential contribution from the
non-resonant $B\rightarrow 4\pi$ and $B\rightarrow \rho\pi\pi$ decays.
For the selection of the \klr\ cut, a systematic error of 
$+2.7$\%\ and  $-2.3$\%\ is measured. We estimate the uncertainty
in the number of $B\bar{B}$ pairs to be  $\pm 1.1$\%. For the efficiency 
calculation with MC we assign a systematic error of  $\pm 0.5$\%.
The reconstruction efficiency ($\varepsilon$) is estimated
with MC simulation  generated with $m_{a_{1}}=1230 \mevm$ and
$\Gamma_{a_1}=400 \mevm$. We vary the   $m_{a_{1}}$
and $\Gamma_{a_1}$ to our measured values and estimate the efficiency
variation, and find that the difference is $-0.5$\%. We thus  add it as a
 systematic error. We also vary the $\Delta E$ range for the 3D fit
to check possible variation of background contribution. The $\Delta E$ is varied from
$-0.10\geve\!<\!\Delta E\!<\!0.10\geve$ to 
$-0.30\geve\!<\!\Delta E\!<\!0.30\geve$.
We find $+4.3$\%\ and  $-1.6$\%\ signal yield variation and assign 
it to the systematic error.
Finally, we add a $+1.8$\%\ systematic error for the fit bias. 
The total  systematic error  is +7.8\%\ and $-8.1$\%.
The branching fraction is measured to be
$\mathcal{B}(\ba1pi)=(48.6\pm 4.1(stat)\pm 3.9(syst) )\times 10^{-6}$.

\begin{table}
\begin{center}
\caption{Systematic errors for the \mbc-\deltaE-$\cos\theta^{}_{{\rm hel}}$ fit.}
\label{tab:syst_meth2}
\renewcommand{\arraystretch}{1.1}
\begin{tabular}{l |c c }
\hline
\hline
{\bf Source of the systematic error} & \multicolumn{2}{ c }{\bf Relative error (\%)} \\\cline{2-3}
            &  {\boldmath $+\sigma$} & {$-$\boldmath$\sigma$} \\
\hline
Track reconstruction efficiency       &  $4.8$    & $-4.8$  \\

PID efficiency                        &  $1.4$    & $-1.4$  \\

Fraction of the non-resonant components &  $0.0$  & $-5.3$ \\

Continuum suppression cut                 &  $2.7$ &  $-2.3$   \\

Signal shape                 &  $1.3$ &  $-0.9$   \\

\a2pi\ background shape                &  $0.4$ &  $-0.3$   \\

Continuum background shape                &  $1.7$ &  $-1.5$   \\

Rare decay background shape                &  $0.4$ &  $-0.4$   \\

Number of \bbbar1\                   &  1.1       &    $-1.1$    \\

MC statistics                         &  $0.5$ &  $-0.5$   \\

$m_{a_{1}}$ parameters in MC                         &  $0.0$ &  $-0.5$   \\

\deltaE\ cut selection                         &  $4.3$ &  $-1.6$   \\

Fit bias                   &  1.8      &    $-0.0$    \\

\hline

Total        &    7.8      &     $-8.1$          \\
\hline
\hline
\end{tabular}
\end{center}
\end{table}


As a validity check of the procedure in the
present analysis we  measure  the branching fraction
of the $B^0\ra D^-\pi^+, D^-\ra K^+\pi^-\pi^-$ decay. 
For this check we perform a  \mbc-\deltaE\ 
unbinned maximum likelihood fit to obtain the signal yield. 
The  branching fraction 
is measured to be  $(2.46\pm 0.05)\times 10^{-3}$, which
is consistent with the world average value of 
$(2.76\pm 0.25)\times 10^{-3}$~\cite{eidleman}.
  
In conclusion, we  have performed a new measurement of the
branching fraction for the  decay, \ba1pi, with  \api\ and \rhopi.
Based on a 253~fb$^{-1}$  data sample, the measured branching fraction is
$\mathcal{B}(\ba1pi)=(48.6\pm 4.1(stat)\pm 3.9(syst)) \times 10^{-6}$, which is consistent with the BaBar measurement~\cite{bab2}. Fitting with an efficiency corrected  relativistic Breit-Wigner, we measure the
parameters of  the $a_{1}(1260)$   to be 
$m_{a_{1}}=1197\pm 34$ \mevm\ and $\Gamma_{a_{1}}=305\pm 43$ \mevm.
These values are consistent with the BaBar measurement~\cite{bab2} and
the world average values~\cite{eidleman}. \\

We thank the KEKB group for the excellent operation of the
accelerator, the KEK cryogenics group for the efficient
operation of the solenoid, and the KEK computer group and
the National Institute of Informatics for valuable computing
and Super-SINET network support. We acknowledge support from
the Ministry of Education, Culture, Sports, Science, and
Technology of Japan and the Japan Society for the Promotion
of Science; the Australian Research Council and the
Australian Department of Education, Science and Training;
the National Science Foundation of China under contract
No.~10175071; the Department of Science and Technology of
India; the BK21 program of the Ministry of Education of
Korea and the CHEP SRC program of the Korea Science and
Engineering Foundation; the Polish State Committee for
Scientific Research under contract No.~2P03B 01324; the
Ministry of Science and Technology of the Russian
Federation; the Ministry of Higher Education, 
Science and Technology of the Republic of Slovenia;  
the Swiss National Science Foundation; the National Science Council and
the Ministry of Education of Taiwan; and the U.S.\
Department of Energy.


\begin{thebibliography}{99}
\bibitem{b0bar}
{ Throughout this paper, the inclusion of  $\bar{B^0}\rightarrow a_{1}^{\pm}(1260)\pi^{\mp}$ decays is implied}.
\end{thebibliography}
\end{document}